# Enhancement of DC-driven flashing corona by dielectric enclosure

Xingxing Wang, Andrei Khomenko, and Alexey Shashurin

Purdue University, School of Aeronautics and Astronautics, West Lafayette, IN 47907, USA

*Abstract* - **In this work, the effect of flashing corona enhancement by introducing Teflon dielectric enclosure in vicinity to the electrode assembly was studied. The discharge operating in air without the dielectric was able to operate within a very narrow voltage range of approximately 200 V. The pulsing frequency was below 1.2 kHz and current peaks were below 14 mA. Increasing the applied voltage onto the positive electrode beyond this range would result in sparks between the electrodes. When the Teflon tube enclosure surrounding the high voltage electrode was used, the window of stable flashing corona operation expanded up to 3-5 kV. The pulsing frequency increased up to 12 kHz and the current peak level increased to approximately 35 mA. Increasing voltage beyond the point with peak pulsing frequency would result in a drop of pulsing frequency until the discharge pulsations stopped completely. The Teflon enclosure was able to enhance the average power deposited into the discharge from 10 to 220 mWatt. In addition, the product gases of the enhanced flashing corona were tested to be mostly ozone with traceable amount of $NO_2$. The discharge used about 150 eV and 1950 eV per one ozone molecule and nitrogen dioxide molecule respectively.**

## I. INTRODUCTION

Atmospheric pressure cold plasmas have been found of great value in the field of bio-engineering, medicine, food processing, etc. due to its ability to produce reactive gas species and radicals with minimal energy consumption and gas heating [1]–[5]. Conventional atmospheric pressure cold plasmas are generated in noble gas, N2, or air by alternating or pulsed HV powers, with a heating of gas by usually less than 50K while electron temperature can be 1-10 eV [1]. Corona discharge is a very widely used method to produce atmospheric pressure cold plasmas [1]. It's been shown to have a great value in the industry for decontamination, pollution control, etc [6]. It is also applicable in flow control over airfoil for being able to increase lift, delay separation, etc. [7].

One type of corona discharge called 'flashing corona' is operating at relatively low voltages where discharge occurs in the intermittent pattern. In a pin-to-plane electrode geometry, with a positive pin electrode separated 31 mm from the plane, flashing corona starts at the voltage of about 5 kV in the form of repetitive current pulses [8], [9]. As voltage increases, the frequency of current pulses peaks at the level of 6.5 kHz. As the voltage keeps increasing, the frequency of current pulses drops back to zero, converting into a 'silent' region called 'pulseless glow corona'. The current level of these pulses is below 1 μA. This kind of discharge behavior is defined as "pre-inception streamers". For pre-inception streamers, as streamer is initiated from the positive electrode, it creates a channel of positive charges that reduce overall field on the anode and prohibit the next streamer from triggering until positive charges are removed from the anode vicinity and the electric field around the anode is restored. This concept was implemented in a DC-driven helium plasma jet device reported recently in works [10], [11].

Utilization of the DC voltage for driving the flashing corona discharge offers significant benefits for practical applications in comparison to use of radio-frequency (RF). This is due to the relief of the electrical insulation requirements, significant reduction of EM radiation, and easy availability of DC high voltage power supplies rather than RF supplies operating in kV range [11]. However, due to its nature of being an auto-oscillator, control of parameters of flashing corona such as current level and pulsing frequency can be troublesome since these parameters are naturally chosen by the system itself without any external control. At the same time, ways to control discharge current and repetition frequency are critical since these determine the 'plasma dose' produced by the plasma generator. In previous work with the helium plasma jet device, several ways of controlling the current level and repetition frequency of flashing corona pulses in relatively narrow range were briefly demonstrated. Specifically, repetition frequency can be increased twice by increasing the helium flow rate from 0.3 to 0.9 LPM, five times by bringing the grounded electrode closer to the anode from 20 cm to 2 cm. In addition, increasing the supplied voltage level from 4.5 to 5 kV resulted in tripling the repetition frequency and doubling the peak current level [10], [11].

Therefore, benefits of DC-driven flashing corona for cold plasma generation were reported previously. However, ways to control and adjust the discharge parameters in DC-driven flashing corona were not studied thoroughly. In this work, we demonstrate that parameters of DC-driven flashing corona can be effectively controlled by introducing dielectric surfaces in proximity to electrode assembly when operating at atmospheric conditions.

## II. EXPERIMENTAL SETUP

A stainless-steel sewing needle with a tip diameter of approximately 200 μm that was fixed onto the end of a high voltage (HV) cable was utilized as the HV central electrode. High voltage up to 20 kV was supplied by a Bertan 225-20 DC HV power supply. A bare copper wire was utilized as the ground electrode. The assemblies were attached to a linear stage separately in order to precisely control the gap distance between the electrodes. When testing the discharge behavior without the dielectric, the grounded wire was directly facing the HV electrode as shown in Fig 1(a). When testing with the Teflon enclosure, the enclosure of Teflon between the two electrodes was introduced into the system by inserting the central electrode cable into a Teflon tube with ID of 1/8" and OD of 1/4". The grounded copper wire was wrapped at the other end of the Teflon tube across its center. A schematic drawing of the plasma device is shown below in Fig 1 (a) and (b). The schematics of the circuitry connection are also shown below in Fig 2. We used a high-power 1 MΩ current-limiting resistor in series with the HV power supply. A Tektronix P6015A HV probe up to 30 kV (Voltage probe 1) was connected to the high voltage electrode to monitor the voltage applied to the discharge (for accurate current measurement, presented below, this probe was unplugged). A non-inductive precision 1% 950 Ω resistor was connected in series on the HV line to measure the current on the high voltage side of the discharge. A special Lecroy high voltage fiber optically isolated probe (HVFO) was used to measure the voltage drop on the resistor, both DC and AC components. On the high voltage side, the current flowing through the circuitry $I$ is express as: $I = \Delta V/R$, where $\Delta V$ is the voltage potential difference on the 950 Ω resistor. In companion to that a Lecroy PP022 probe (Voltage probe 2) was connected to the grounded copper wire, which was coupled to ground by another precision 1% 950 Ω resistor. Additionally, we investigated an AC component of the HV side current by a Bergoz fast current transformer (not shown in Figure 2). Thus, we had evaluated three different readings – two for the HV side and one for Ground side – and found them in good agreement between each other for all working regimes. Further, only Ground side current shunt was used for current measurement. The oscilloscope used in this experiment was Lecroy HDO 800 with bandwidth of 3 GHz.

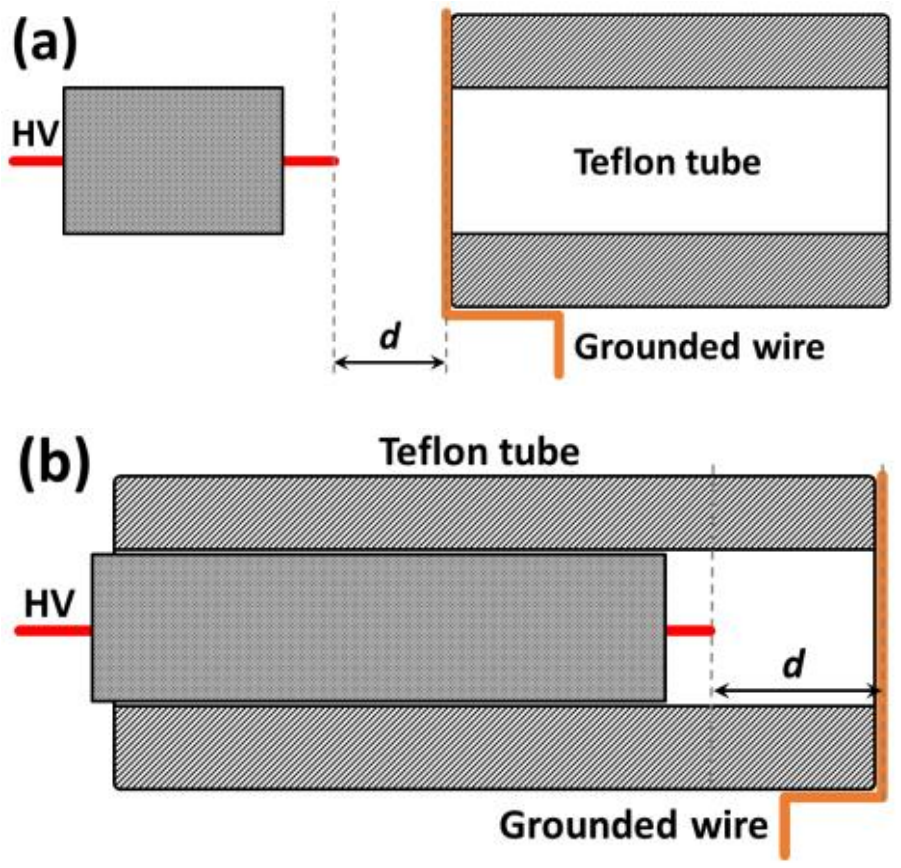

**Fig 1. Experimental setup for the case without Teflon enclosure (a) and with Teflon enclosure (b).**

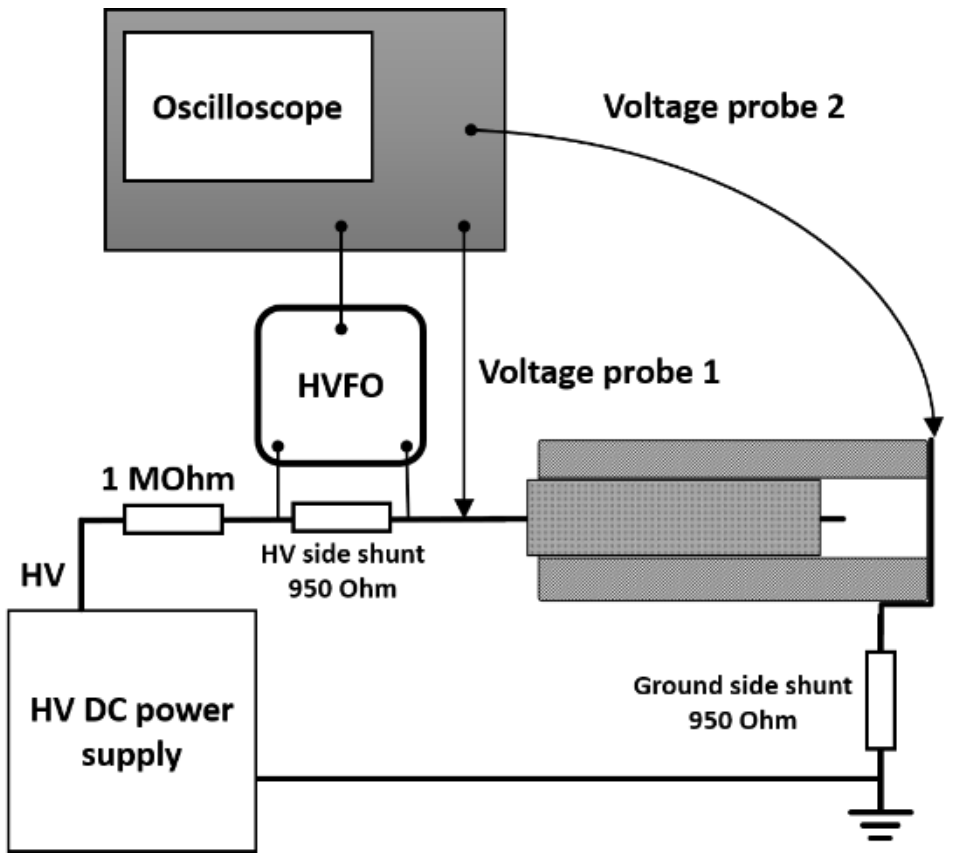

**Fig 2. Electric circuitry used in the experiment.**

In order to measure the reactive gases produced by the flashing corona, the Teflon tube was placed in an acrylic enclosure where a constant air stream blows on the discharge, as shown in Figure 3. The downstream flow was directed to gas analyzers. For nitric oxide and nitrogen dioxide we utilized a Thermo 42C high level NO-$NO_2$-$NO_x$ analyzer with a measuring principle based on chemiluminescence, and for ozone a custom-built device based on UV absorption at 254 nm was used.

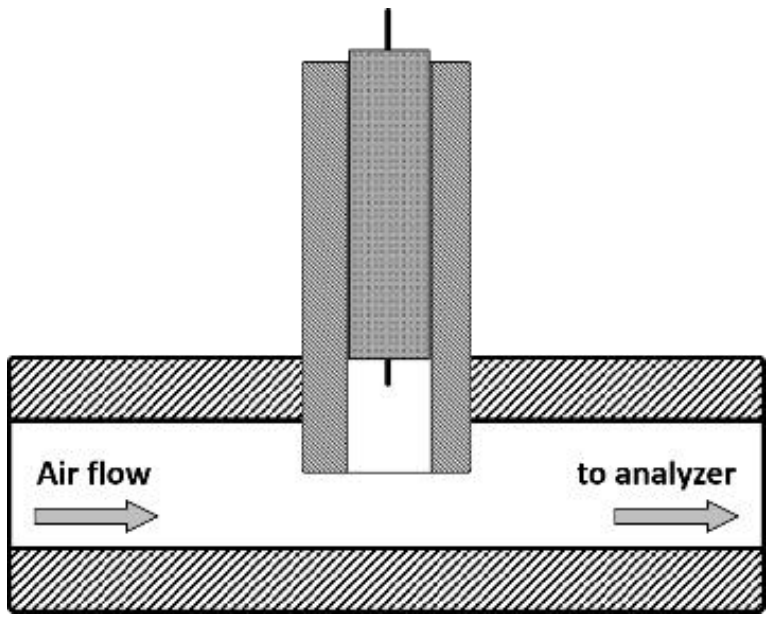

**Fig 3. Schematics of the system utilized for testing reactive gases produced by the plasma source.**

## III. EXPERIMENTAL RESULTS AND DISCUSSIONS

At a fixed gap distance between electrodes *d*, when increasing the voltage *V*, the current pulses would start to initiate in the form of irregular pulsations. As *V* increases, the current pulse became more stable and repeatable. An example of the continuous current pulses is shown below in Fig 4(a). A close look of a single current pulse is also shown in Fig 4(b). One can see from Fig 4(b) that individual discharge event consists of two stages: at the beginning stage, the current peak of about 14 mA and duration of about 100 ns was generated; while at the later stage, current pulse of about 100 μA and duration 10s of μs was observed. The shape and duration of each individual pulse in all conditions tested in this work were very repeatable with variations <10%.

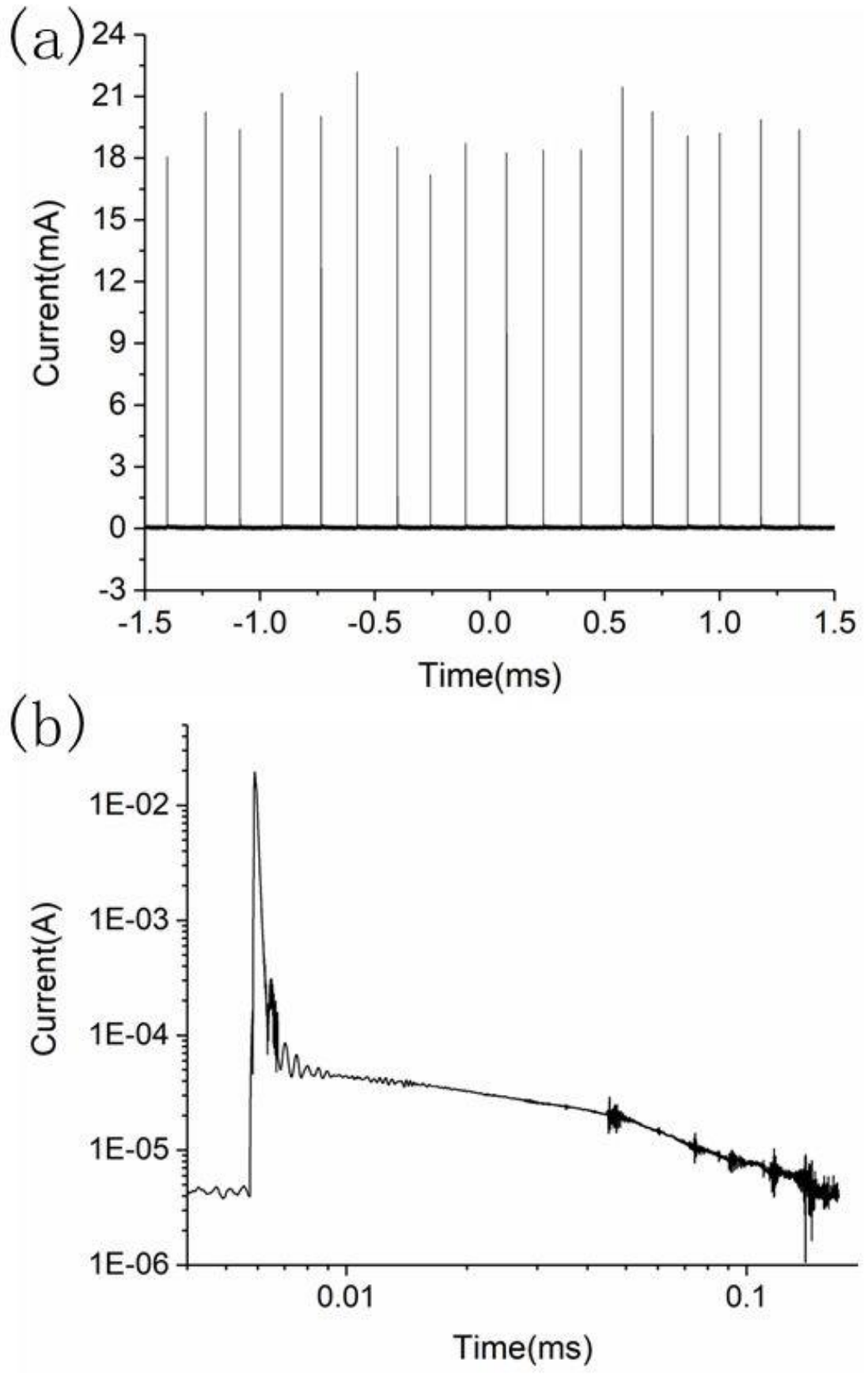

**Fig 4. Temporal evolution of current of DC-driven flashing corona without dielectric. (a) Series of current pulses, (b) Typical appearance of the individual current pulse.**

The gap distance *d* was adjusted between 4 and 12 mm with the increment of 1 mm. For the case of electrode assembly without the Teflon enclosure shown in the Fig 5(a), the dependencies of flashing corona pulsing frequency and peak current value vs. *V* at different gap distances are shown below as Fig 5(a) and Fig 5(b). It was observed that starting from the moment when continuous pulses started to operate, as *V* increases, both pulsing frequency and peak current increases rapidly and quickly reaches the level that leads to sparks. It can be seen that the voltage range of continuous pulsing operation was very narrow, approximately 200 V at each gap distance. Within that range, when *d* =4 mm, frequency increased from $f_{min}$ =0.01 kHz to $f_{max}$ =1.2 kHz when voltage increased from 4.7 to 4.9 kV. As the gap distance increases, the frequency range shrinks by both having a greater minimum value $f_{min}$ and a smaller maximum value $f_{max}$. At *d* = 12 mm, the frequency increased from 0.5 kHz to 0.58 kHz when voltage increased from 16.6 to 16.9 kV. On the other hand, when *d* increases, the minimum value of current pulses $I_{peak}$ increases from 7.5 to 13 mA. However, the maximum value of $I_{peak}$ saturates at the level of approximately 14 mA starting from *d* = 5 mm.

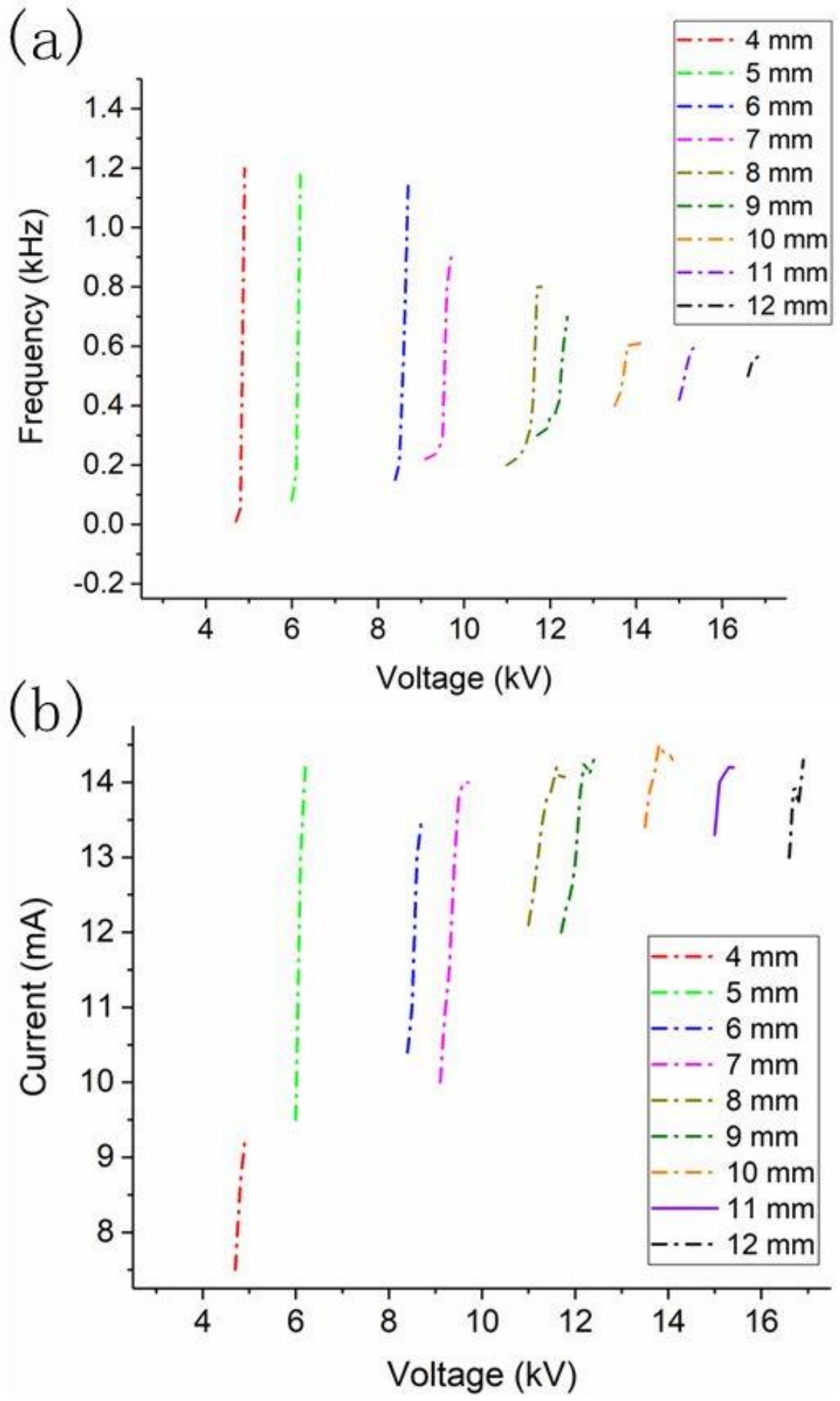

**Fig 5. Frequency of current pulses vs. DC voltage at different gap distances without Teflon (a). Peak value of current pulses vs DC voltage at different gap distances without Teflon (b).**

For the case when the Teflon enclosure was introduced into the system as shown in the Fig 1(b), the frequency and current peak value of stable flashing corona pulses were measured as shown below in Fig 6(a) and (b). A bell-shaped distribution of pulsing frequency vs. driving DC voltage was observed for all gap distances as shown in Fig 6(a), indicating that as voltage increases, the device was able to reach the 'silent' pulseless region after reaching the maximum pulsing frequency. For $d$ = 5mm, the pulsing frequency increases from zero up to 12 kHz at $V$ = 6.5 kV, then drops back to zero at $V$ = 7.8 kV. One can also see that, as $d$ increases, $f_{max}$ drops from 12 kHz at $d$ = 5mm, down to 2.2 kHz at $d$ = 11mm. In addition, as $d$ increases, the range of operating voltage expands from 3 to 5 kV. On the other hand, the peak value of current pulses shows a similar trend as without the Teflon enclosure, as shown in Fig 4(b) and Fig 6(b). For each $d$, as voltage increases, $I_{peak}$ increases. The minimum value of $I_{peak}$ also increases with $d$, from 12 mA at $d$ = 5 mm up to 30 mA at $d$ >10 mm. Saturation of $I_{peak}$ was also observed. The saturation level of value of current peaks was about 35 mA and could be achieved when $d \geq 5$ mm.

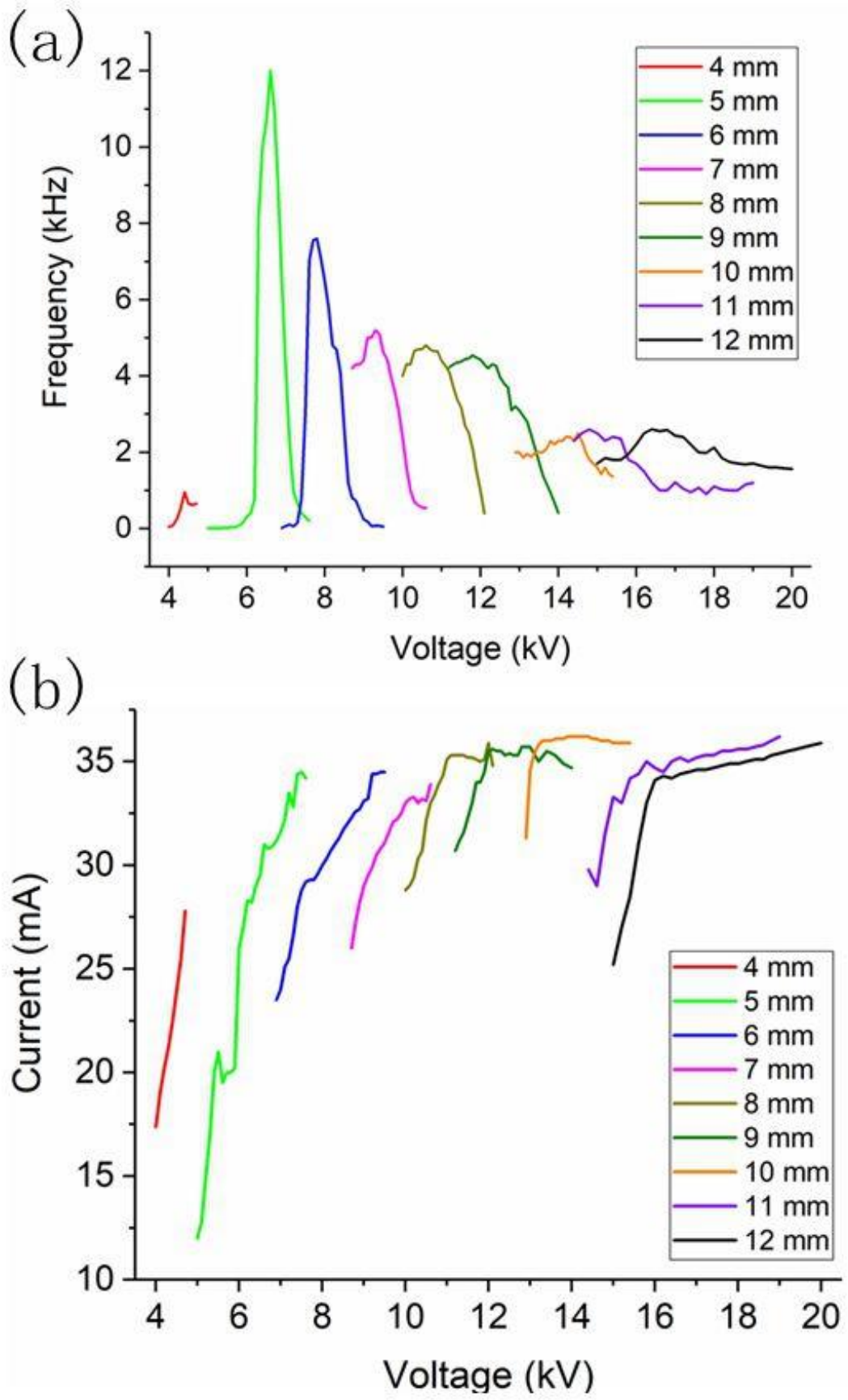

**Fig 6. Frequency of current pulses vs. DC voltage at different gap distances with Teflon enclosure (a). Peak value of current pulses vs DC voltage at different gap distances with Teflon enclosure (b).**

By comparing the operation of this plasma source with and without the Teflon enclosure, it can be concluded that a Teflon tube enclosure can greatly stabilize the current pulsations of flashing corona source and expand the operational voltage range. It also enhances the pulsations by both increasing the pulsing frequency by one order of magnitude and peak current value. In addition, it enables a 'silence' region between the continuous current pulsing and strong sparks, which is important for practical applications where stable operation without sparks is desirable. The average power consumption of the flashing corona in both cases is calculated as $P = f \int_{\text{one pulse}} I\, V dt$. Fig shows a comparison of average discharge power with and without the Teflon enclosure for $d$ = 5 mm. One can see that the operational voltage range is much wider with the application of a Teflon enclosure. In addition, the Teflon enclosure significantly increases the average discharge power from 10 mW up to 220 mW.

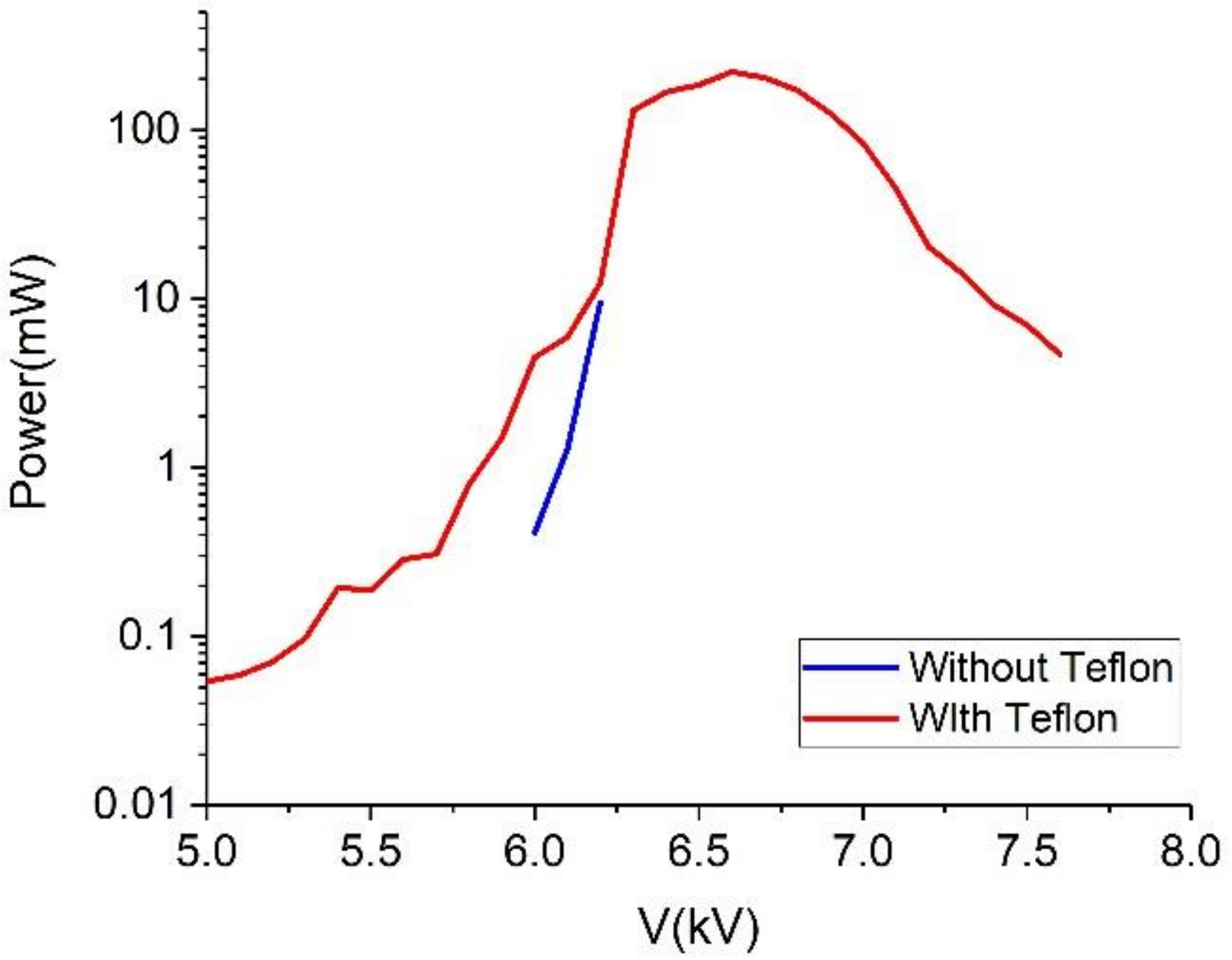

**Fig 7. Comparison of the average discharge power with and without Teflon enclosure for 5 mm gap distance.**

The effect of enhancement of flashing corona using a dielectric enclosure can be potentially explained by the presence and accumulation of nitrogen metastable molecules $N_2^*$ created during the discharge. Indeed, the Teflon enclosure tube inhibits these metastable molecules from diffusing away from the electrodes thus increasing the local concentration of metastable molecules in the vicinity of the electrodes [12]. Higher concentration of the metastable molecules can cause production of more seed electrons due to the associative ionization mechanism [13]. These additional seed electrons thus enhance the avalanches developing in the front of the streamer head and cause streamer inception at lower electric fields on the anode tip.

Therefore, the enhanced flashing corona discharge looks promising for nitrogen oxides and ozone generation because of higher energies deposited to the discharge. For normal corona type discharges, it’s a well-known fact that they are used for ozone generation. Generally, researchers use air or oxygen filled short gaps where one of the electrodes is covered with an insulating layer in order to prevent streamer-arc transition, so-called silent discharge [14]. Thus, we expected significant ozone production in the enhanced flashing corona regime and following experiments were conducted in order to confirm these speculations. The discharge had been placed in the additional acrylic case to cover the airstream as shown in Fig 3(b). It has to be noted, that stable operational range has slightly changed compared to that then the discharge was fully exposed to the stagnant ambient air. For gas analysis we picked an operational point where the gap was 10 mm, the DC voltage was 10-10.5 kV, and the resulting pulse repetition frequency was about 6 kHz. An average discharge consuming power in such regime is close to 220 mW or $1.4 \cdot 10^{18}$ eV/s.

By varying air flow rate, ozone, nitric oxide and nitrogen dioxide concentrations were measured, as presented in Fig 8. The nitric oxide graph is not shown because its amount was always below the measuring limit of our equipment. Absence of nitric oxide has been predicted after we found a significant amount of ozone in the corona’s exhaust. Indeed, in an ozone rich mixture NO quickly converts to $NO_2$ according to the following reaction [15]: $NO + O_3 \rightarrow NO_2 + O_2$.

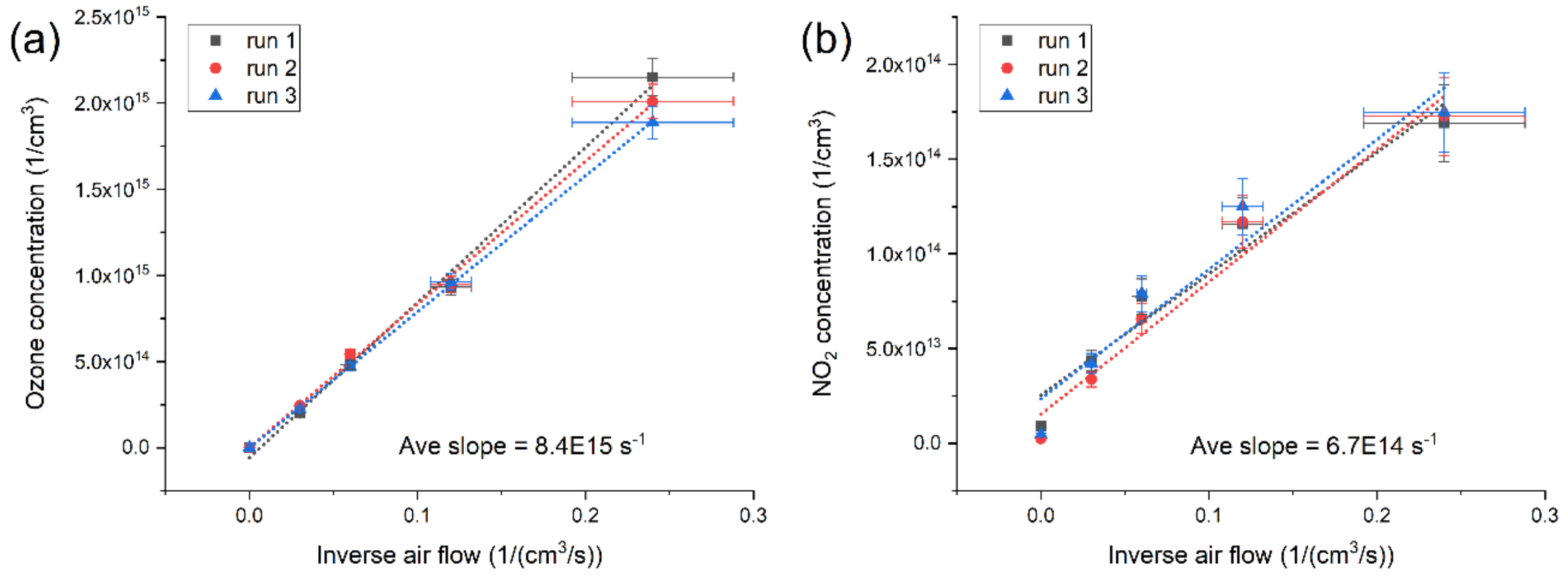

**Fig 8. Ozone (a) and nitrogen dioxide (b) concentrations with respect to inverse air flow. The points with zero inverse air flow (or infinite air flow) correspond to plasma-off case.**

According to the Fig 8(a), ozone concentration increased linearly with inverse air flow which is indicative of constant ozone production regardless of the air flow rate, but for $NO_2$ we see slightly non-linear behavior – low $NO_2$ than it should be at 0.5 and 0.25 l/min – it indicates the discharge does not have enough fresh air to successfully evacuate all the chemical products from the space around it at these pretty low flow rates.

If one approximates the line slope in the Fig 8, a molecule production rate in one second can be found. By dividing the average discharge power to the production rate, the energy cost per one molecule may be determined. The enhanced corona discharge has about 150 and 1950 eV/molecule production efficiencies for ozone and $NO_2$ respectively.

## IV. CONCLUSION

In this work a DC-driven plasma device that utilizes flashing corona discharges was studied, and the effect of the Teflon enclosure between electrodes in order to control the characteristics of the discharge was demonstrated. It was shown that a Teflon enclosure can increase the amplitude and frequency of current pulses. At a gap distance of 5 mm, the average discharge power was increased from 10 to 220 mWatt due to the Teflon enclosure. In addition, use of the Teflon enclosure causes a significantly more stable operation of the discharge. The product reactive gases of the enhanced flashing corona were tested to be mostly ozone with traceable amount of $NO_2$. The discharge uses about 150 eV and 1950 eV per one ozone molecule and nitrogen dioxide molecule, respectively.

## V. REFERENCES

[1] A. Fridman, *Plasma chemistry*. Cambridge: Cambridge University Press, 2008.

[2] M. Laroussi, “The Biomedical Applications of Plasma: A Brief History of the Development of a New Field of Research,” *IEEE Trans. Plasma Sci.*, vol. 36, no. 4, pp. 1612–1614, Aug. 2008.

[3] M. Keidar *et al.*, “Cold plasma selectivity and the possibility of a paradigm shift in cancer therapy,” *Br. J. Cancer*, vol. 105, no. 9, pp. 1295–1301, Oct. 2011.

[4] A. Shashurin, M. Keidar, S. Bronnikov, R. A. Jurjus, and M. A. Stepp, “Living tissue under treatment of cold plasma atmospheric jet,” *Appl. Phys. Lett.*, vol. 93, no. 18, p. 181501, Nov. 2008.

[5] G. E. Morfill and J. L. Zimmermann, “Plasma Health Care - Old Problems, New Solutions,” *Contrib. to*

*Plasma Phys.*, vol. 52, no. 7, pp. 655–663, Aug. 2012.

[6] Y.-H. Lee *et al.*, “Application of Pulsed Corona Induced Plasma Chemical Process to an Industrial Incinerator,” *Environ. Sci. Technol.*, vol. 37, no. 11, pp. 2563–2567, Jun. 2003.

[7] T. C. Corke, C. L. Enloe, and S. P. Wilkinson, “Dielectric Barrier Discharge Plasma Actuators for Flow Control,” *Annu. Rev. Fluid Mech.*, vol. 42, no. 1, pp. 505–529, Jan. 2010.

[8] Y. P. Raizer, *Gas Discharge Physics*, 1st ed. Springer-Verlag Berlin Heidelberg, 1991.

[9] M. Goldman and A. Goldman, “Corona Discharges,” in *Gaseous Electronics*, Elsevier, 1978, pp. 219–290.

[10] X. Wang and A. Shashurin, “DC-driven plasma gun: self-oscillatory operation mode of atmospheric–pressure helium plasma jet comprised of repetitive streamer breakdowns,” *Plasma Sources Sci. Technol.*, vol. 26, no. 2, p. 02LT02, Jan. 2017.

[11] X. Wang and A. Shashurin, “Study of atmospheric pressure plasma jet parameters generated by DC voltage driven cold plasma source,” *J. Appl. Phys.*, vol. 122, no. 6, p. 063301, Aug. 2017.

[12] G. Hartmann and I. Gallimberti, “The influence of metastable molecules on the streamer progression,” *J. Phys. D. Appl. Phys.*, vol. 8, no. 6, pp. 670–680, Apr. 1975.

[13] N. A. Popov, “Associative ionization reactions involving excited atoms in nitrogen plasma,” *Plasma Phys. Reports*, vol. 35, no. 5, pp. 436–449, May 2009.

[14] M. Goldman, A. Goldman, and R. S. Sigmond, “The corona discharge, its properties and specific uses,” *Pure Appl. Chem.*, vol. 57, no. 9, pp. 1353–1362, Jan. 1985.

[15] A. C. Gentile and M. J. Kushner, “Reaction chemistry and optimization of plasma remediation of NxOy from gas streams,” *J. Appl. Phys.*, vol. 78, no. 3, pp. 2074–2085, Aug. 1995.